\documentclass[aps,prb,twocolumn]{revtex4-1}
\usepackage{amssymb}
\usepackage{graphicx}

\begin{document}
\author{Ashiwini Balodhi and Yogesh Singh}
\affiliation{Indian Institute of Science Education and Research Mohali, Sector 81, S. A. S. Nagar, Manauli PO 140306, India}

\date{\today}

\title{Synthesis and pressure and field dependent magnetic properties of the Kagome-bilayer spin liquid Ca$_{10}$Cr$_7$O$_{28}$}

\begin{abstract}
We report synthesis of polycrystalline samples of the recently discovered spin liquid material Ca$_{10}$Cr$_7$O$_{28}$ and present measurements of the ambient and high pressure magnetic susceptibility $\chi$ versus temperature $T$, magnetization $M$ versus magnetic field $H$ at various $T$, and heat capacity $C$ versus $T$ at various $H$.  The ambient pressure magnetic measurements indicate the presence of both ferromagnetic and antiferromagnetic exchange interactions with dominant ferromagnetic interactions and with the largest magnetic energy scale $\sim 10$~K\@.  The $\chi(T)$ measurements under externally applied pressure of up to $P \approx 1$~GPa indicate the robust nature of the spin-liquid state despite relative increase in the ferromagnetic exchanges.  $C(T)$ shows a broad anomaly at $T\approx 2.5$~K which moves to higher temperatures in a magnetic field.  The evolution of the low temperature $C(T,H)$ and the magnetic entropy is consistent with frustrated magnetism in Ca$_{10}$Cr$_7$O$_{28}$. 
 
\end{abstract}
\maketitle

\section{Introduction}
Most local moment magnets undergo a transition from a high temperature paramagnetic state to a magnetically long range ordered state below some critical temperature.  The ordered state is like a solid while the paramagnetic state is like the gaseous state.  The state analogous to a liquid is elusive in magnets.  That is because most magnetic solids have a unique state with the lowest energy and are able to freeze into that solid-like long range ordered state at sufficiently low temperatures.  If one is able to suppress this tendency to order, one could attain a liquid like state of spins where they would be strongly entangled and yet dynamically fluctuating down to $T = 0$.  This can be achieved by enhancing quantum fluctuations which can melt the magnetic solid.  One way of constructing such a quantum spin-liquid (QSL) is by arranging the magnetic moments on low dimensional or geometrically frustrated lattices.  
Indeed, QSL's were first demonstrated to exist for quasi-one-dimensional spin chains like KCuF$_3$ and Sr$_2$CuO$_3$ (see Refs.~\onlinecite{Yamashita-review2000, Lemmens2003} for experimental reviews
).   

The quest for spin liquid realizations in higher dimensions has led to a flurry of activity in the last two decades resulting in the discovery of quite a few candidate spin liquid materials \cite{Mila2000, Balents2010}.  The best established candidates are the quasi-two-dimensional Kagome lattice quantum magnet Herbertsmithite ZnCu$_3$(OH)$_6$Cl$_2$,\cite{Helton2007,Han2012} the triangular lattice organic magnets $\kappa$--(BEDT-TTF)$_2$Cu$_2$(CN)$_3$ (Refs.~\onlinecite{S-Yamashita2008, M-Yamashita2009}) and EtMe$_3$Sb[Pd(dmit)$_2$]$_2$,\cite{M-Yamashita2010} and the recently discovered Yb based triangular lattice magnet YbMgGaO$_4$ \cite{Li2015}.  There are some other materials which showed promise either because of their geometrically frustrated lattice, like the 3-dimensional hyper-kagome iridate Na$_4$Ir$_3$O$_8$, or because of novel frustration mechanisms like the Kitaev QSL candidates $A_2$IrO$_3$.  Both these material families however, showed freezing or magnetic order at low temperatures \cite{Singh2013, Shockley2015, Singh2010, Singh2012}.   

Recently a new quasi two-dimensional quantum magnet Ca$_{10}$Cr$_7$O$_{28}$ has been discovered with a Kagome bilayer structure \cite{Balz2016}.  Using bulk measurements like magnetic susceptibility and heat capacity, and microscopic measurements like $\mu$SR and neutron scattering this material has been reported to show all the expected signatures of a gapless quantum spin liquid (QSL) \cite{Balz2016}.  The spin-liquid state in Ca$_{10}$Cr$_7$O$_{28}$ has been shown to develop from a novel frustration mechanism where competing ferro- and anti-ferromagnetic exchange interactions within a Kagome-bilayer suppress the possibility of long-ranged magnetic order.  How this spin-liquid state evolves if the balance between the competing interactions is disturbed by external perturbations like hydrostatic pressure or magnetic field, is the question we address in this work. 

Here we report synthesis and structure of polycrystalline samples of the recently discovered QSL Ca$_{10}$Cr$_7$O$_{28}$.  We present a detailed study of ambient and high pressure magnetic susceptibility $\chi$ versus temperature $T$, magnetization $M$ versus magnetic field $H$ at various $T$, and heat capacity $C$ versus $T$ at various $H$.

\section{Experimental Details}
Polycrystalline samples of Ca$_{10}$Cr$_7$O$_{28}$ were prepared by solid state synthesis. The starting materials CaCO$_3$ (99.99\%, Alfa Aesar) and Cr$_2$O$_3$ (99.99\%, Alfa Aesar) were taken to make the Ca:Cr ratio $10.5:7$ and mixed thoroughly in an agate mortar and pelletized.  The pellet was placed in a covered Al$_2$O$_3$ crucible, heated in air at $750~^o$C for $24$~hrs for calcination and then heated to $1000~^o$C for $48$~hrs, and then quenched in Argon to room temperature. After the initial heat treatment, the material was reground and pressed into a pellet and given two heat treatments of $24$~h each at $1100~^o$C, with an intermediate grinding and pelletizing step.  The pellet was always brought to room temperature by quenching in Argon.  Hard well sintered pellets were obtained which are dark green in color. Powder x-ray diffraction (PXRD) patterns were obtained at room temperature using a Rigaku diffractometer with Cu K$\alpha$ radiation, in the $2\theta$ range from $10^o$ to $90^o$ with a step size of $0.02^o$. Intensity data were accumulated for $5$~s/step. The ambient pressure magnetization $M$ versus temperature $T$ and magnetic field $H$ were measured using a VSM option of a Quantum Design physical property measurement system (QD-PPMS), and the heat capacity $C$ as a function of $T$ and $H$ was measured using a QD-PPMS.  The high pressure $\chi$ versus $T$ for pressures upto $P \approx 1$~GPa were measured in a SQUID magnetometer for Cryogenics Limited (CL-SQUID) using a piston-clamp based high pressure cell.

\begin{figure}[h]   
\includegraphics[width= 3.8 in]{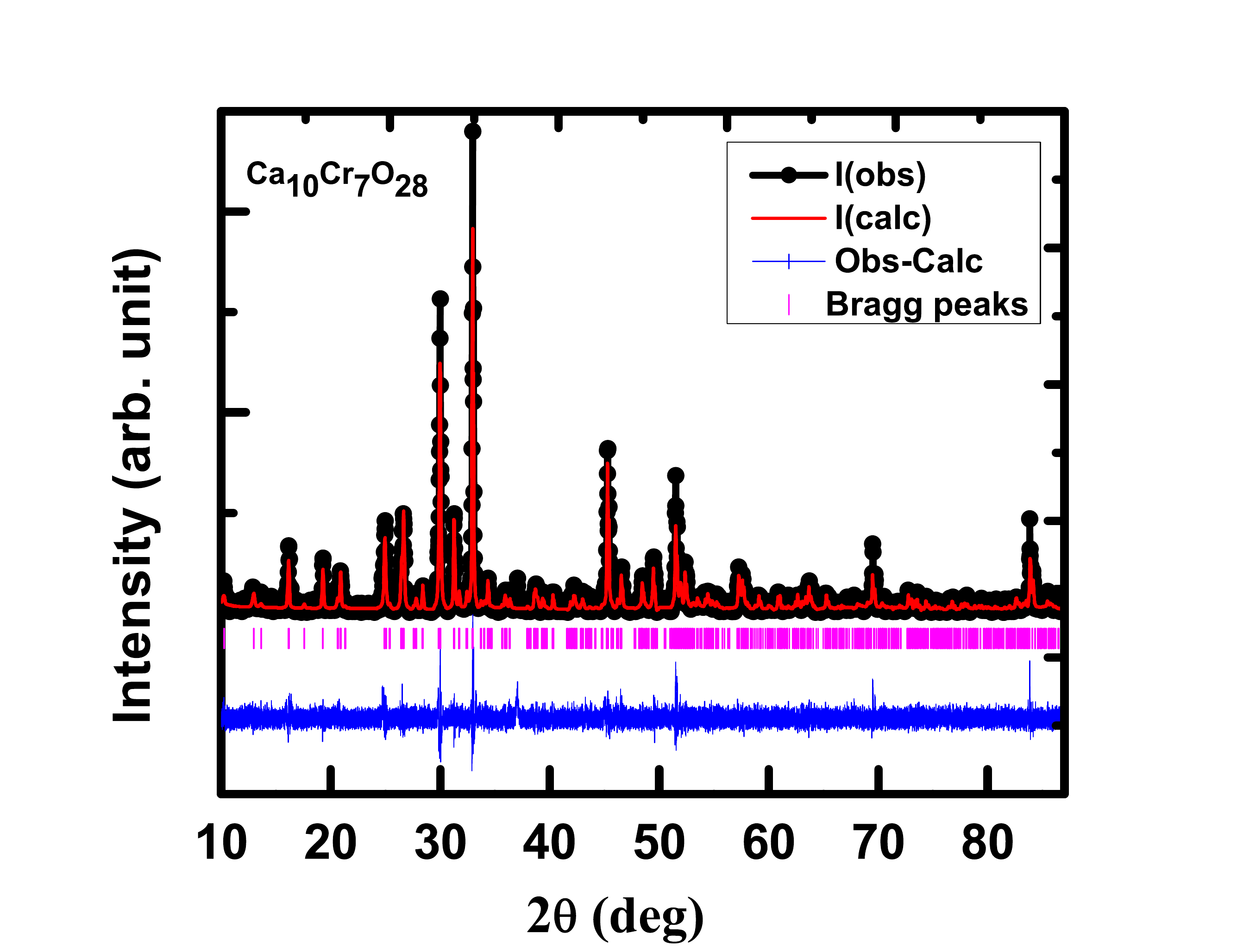}    
\caption{(Color online) Rietveld refinement of powder x-ray diffraction data for Ca$_{10}$Cr$_7$O$_{28}$. The solid circles represent the observed data, the solid lines through the data represent the fitted pattern, the vertical bars represent the peak positions, and the solid curve below the vertical bars is the difference between the observed and the fitted patterns.  
\label{Fig-xrd}}
\end{figure} 

\begin{figure}[h]   
\includegraphics[width= 3.5 in]{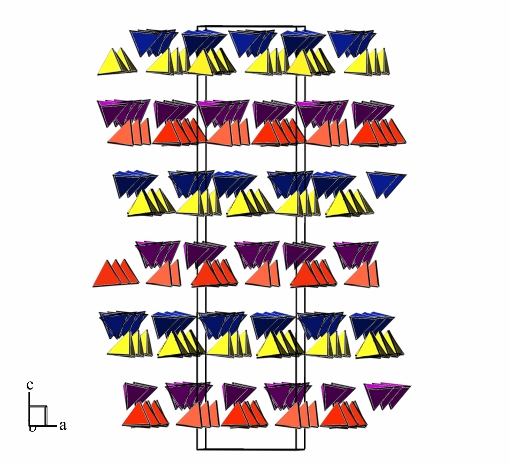}    
\includegraphics[width= 3 in]{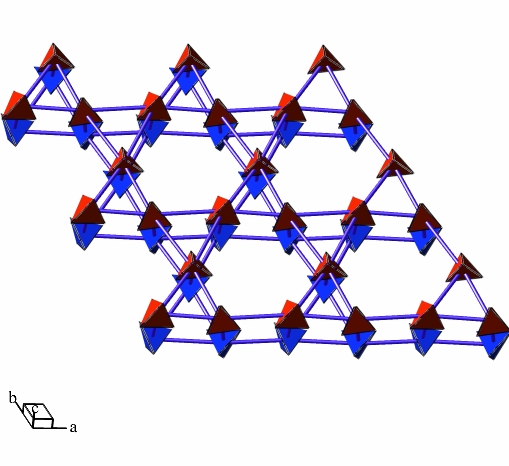}    
\caption{(Color online) Structure of Ca$_{10}$Cr$_7$O$_{28}$ viewed perpendicular (top panel) to the $c$-axis.  The $4$ inequivalent CrO$_4$ tetrahedra are shown in different colors.  Planes formed by these CrO$_4$ tetrahedra are stacked along the $c$-axis.  A view approximately down (bottom panel) the $c$-axis showing the bi-layer Kagome network of Cr ions.  
\label{Fig-structure}}
\end{figure}

\section{Results}
\subsection{Structure}
The powder x-ray diffraction (PXRD) data is shown in Fig.~\ref{Fig-xrd}.  The PXRD data could be refined starting with the structural model recently reported for Ca$_{10}$Cr$_7$O$_{28}$~\cite{Balz2016, Balz2016a}.  The Rietveld refinement results are also plotted in Fig.~\ref{Fig-xrd}.  The lattice parameters obtained from the fit are $a = b = 10.7885(3)$~\AA, $c = 38.163(1)$~\AA, $\alpha = \beta = 90^o$, $\gamma = 120^o$, and the cell volume $V = 3846.79(20)$~\AA$^3$.  The Wyckoff atomic positions, site occupancies, and thermal parameters obtained from the fit are given in Table~\ref{Structure}.  These parameters match well with recently reported data on single crystals and confirm the single phase synthesis of polycrystalline Ca$_{10}$Cr$_7$O$_{28}$. \citep{Balz2016,Balz2016a}  

Six out of the seven Cr ions per formula unit are in the Cr$^{5+}$ valence state while one is in the Cr$^{6+}$ valence state \cite{Balz2016}.  The Cr$^{5+}$ ions are magnetic with $S = 1/2$ while the Cr$^{6+}$ ion is non-magnetic.  The magnetic results discussed below are consistent with this distribution.  Figure~\ref{Fig-structure} shows the arrangements of the magnetic Cr ions.  These ions sit in oxygen tetrahedra.  There are $4$ inequivalent tetrahedra shown as the $4$ different colors in Fig.~\ref{Fig-structure}.  The structure is made up of layers of these different tetrahedra alternating along the $c$-axis as shown in the top panel of Fig.~\ref{Fig-structure}.  Within the layers the tetrahedra are arranged on a Kagome lattice as can be seen in the lower panel in Fig.~\ref{Fig-structure}.  The magnetic connectivity between the Kagome layers is such that the blue and orange Kagome layers interact with each other but are isolated from the other layers.  Similarly, the yellow and purple Kagome layers interact but are isolated from the others.  Thus, magnetically, Ca$_{10}$Cr$_7$O$_{28}$ can be viewed as a layered bi-layer Kagome system.

\begin{table}[h!]
\centering
\caption{Atomic parameters obtained by refining x-ray powder diffraction for Ca$_{10}$Cr$_7$O$_{28}$ with a space group 167, $R3c$. The lattice constants are $a = b = 10.788(5)$~\AA, $c = 38.163$~\AA, $\alpha = \beta = 90$ and $\gamma = 120^o$ }
  \label{tab:table1}
  \begin{tabular} { |c |c| c| c| c| c| c |  }
\hline    	
    	Atom  &  Wyck   & x & y & z & Occ & B (\AA)  \\
    \hline
   Ca1& 36f& 0.292(4) & 0.165(8) & -0.070(6) &  0.97 & 0.0058\\
   Ca2& 36f& 0.182 & -0.204(3) & -0.008(5) &  0.96 & 0.0071\\
   Ca3& 36f& 0.374 & 0.157(7) & 0.022 & 1 & 0.0086\\
   Ca4& 12c& 0.666(7) & 0.333 & 0.087 & 1 & 0.0099\\
   Cr1& 36f& 0.316(2) & 0.154(5) & 0.124 & 1 & 0.0012\\
   Cr2& 36f& 0.164(7) & -0.142 & -0.107 &  0.95 & 0.0143\\
   Cr3A& 12c& 0.000 & 0.000 & -0.018(4) &  0.64(5) & 0.0031\\
   Cr4B& 12c& 0.000 & 0.000 & -0.007(5) &  0.33 & 0.0064\\
   O3A& 12c& 0.000 & 0.000 & -0.023(2)&  0.65(9) & 0.001\\
   O3B& 12c& 0.000 & 0.000 & -0.364&  0.0256(9) & 0.0900\\
   O1& 36f& 0.300(9) & 0.189(8) & 0.054(9)&  0.84(8)& 0.0192\\
   O2& 36f& 0.209 & 0.173(4) & 0.182(5)&  1.000& 0.0110\\
   O3& 36f& 0.116(9) & -0.0331 & 0.130(7)&  1.000& 0.0395\\
   O4& 36f& 0.367 & 0.147(6) & 0.131&  1.000& 0.0008\\
   O5& 36f& 0.1239 & -0.155(7) & -0.062&  0.95(4)& 0.033\\
   O6& 36f& 0.231(7) & -0.267(6) & -0.124(4)&  1.000& 0.0081\\
   O8& 36f& -0.010(3) & -0.227(7) & -0.118(4)&  0.86(5)& 0.0094\\
   O9& 36f& -0.106 & 0.011(5) & 0.006(6)&  0.91(3)& 0.0131\\ \hline
\label{Structure}
  
  \end{tabular}
\end{table}

\begin{figure}[h]   
\includegraphics[width= 3 in]{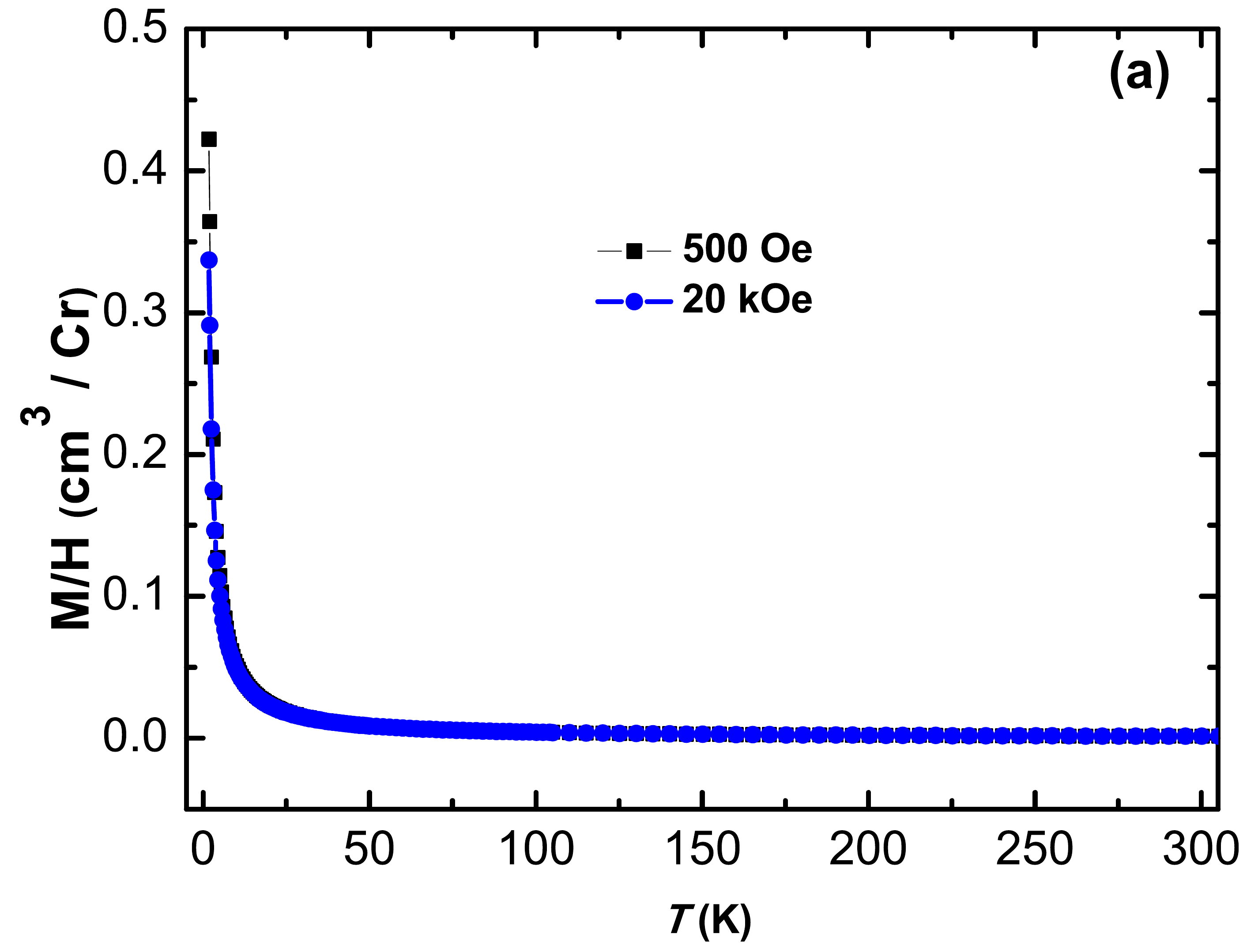}    
\includegraphics[width= 3 in]{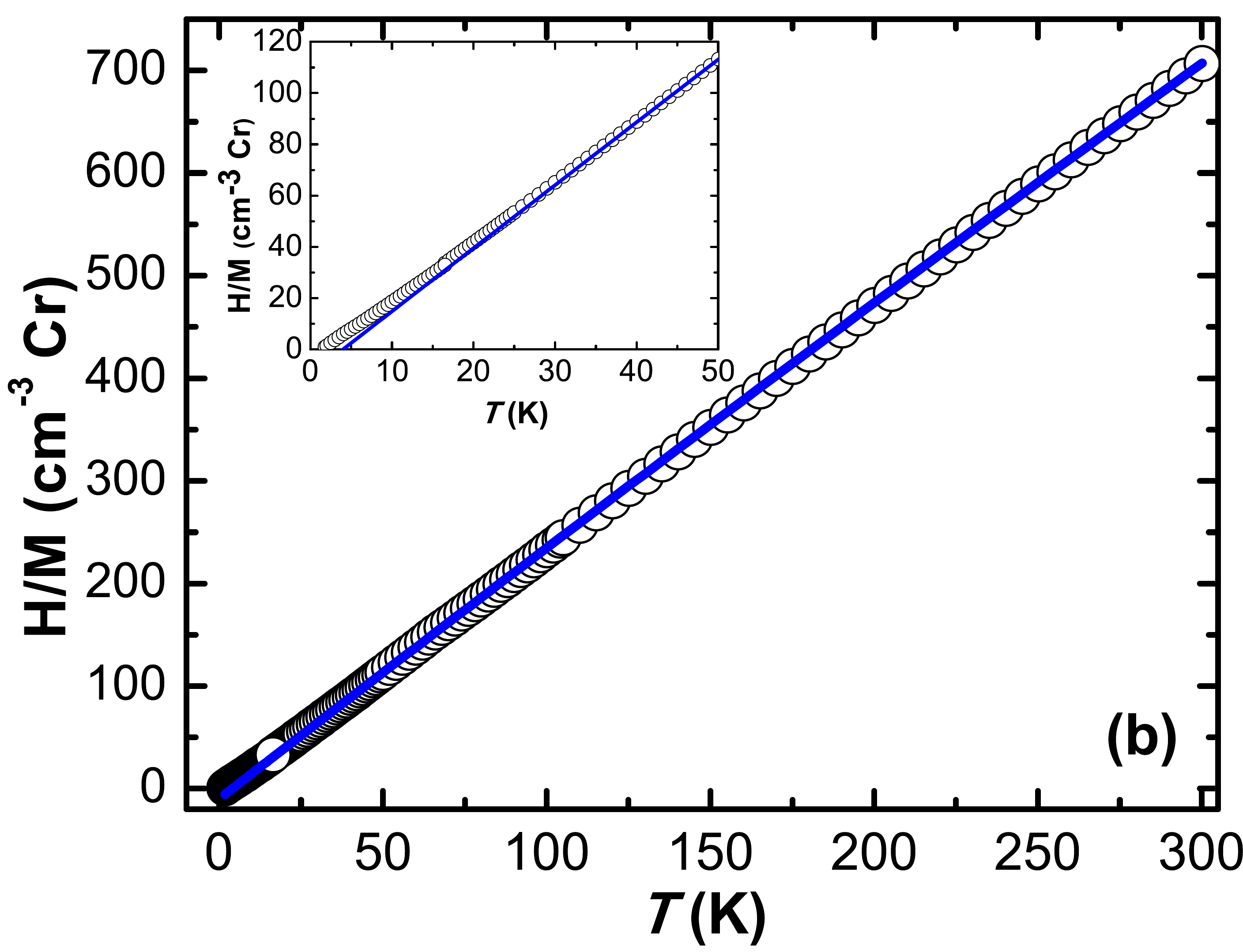}    
\caption{(Color online) (a) Magnetization $M$ divided by magnetic field $H$ versus temperature $T$ for Ca$_{10}$Cr$_7$O$_{28}$ measured in $H = 1$ and $20$~kOe.  (b) $H/M$ versus $T$ measured at $H = 20$~kOe.  The solid curve through the data is a fit by the Curie-Weiss expression.  The inset shows the $H/M$ data below $T = 50$~K to highlight the deviation of the data from the Curie-Weiss fit. The data is presented per Cr in the formula unit (which is 7).   (see text for details). 
\label{Fig-chi}}
\end{figure}

\begin{figure}[h]   
\includegraphics[width= 3 in]{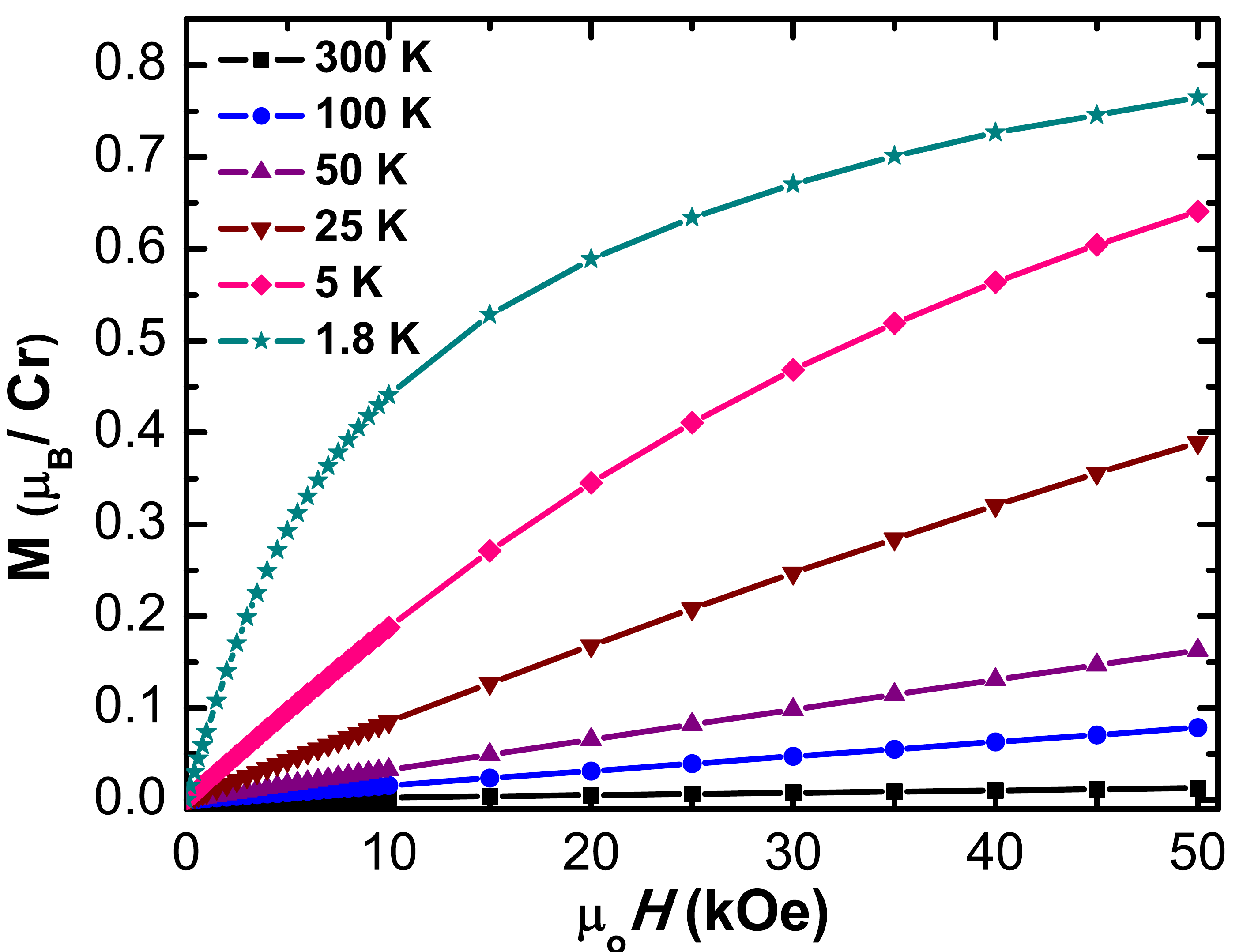}    
\caption{(Color online) Magnetization $M$ versus magnetic field $H$ for Ca$_{10}$Cr$_7$O$_{28}$ measured at various temperatures $T$.  The data is presented per Cr in the formula unit (which is 7). 
\label{Fig-MH}}
\end{figure}

\subsection{Ambient Pressure Magnetic Results}
Figure~\ref{Fig-chi}~(a) shows the magnetic susceptibility $\chi = M/H$ versus temperature $T$ of Ca$_{10}$Cr$_7$O$_{28}$ measured in a magnetic field $H = 1$ and $20$~kOe.  Curie-Weiss like local moment paramagnetism is clearly evident.  At low temperatures $\chi$ increases rapidly and reaches values which are quite large and similar to values seen in ferromagnetic materials.  However, no anomaly signalling long ranged magnetic order is observed down to $T = 1.8$~K\@.   

The $1/\chi(T)$ data at high temperatures $T \geq 200$~K were fit by the Curie-Weiss expression $\chi = \chi_0 + {C \over {T-\theta}}$, where $\chi_0$ is a $T$ independent contribution, $C$ is the Curie constant, and $\theta$ is the Weiss temperature.  The $1/\chi(T)$ data and the fit are shown in Fig.~\ref{Fig-chi}~(b).  The fit gave the values $\chi_0 = -5.2(4) \times 10^{-5}$~cm$^3$/Cr, $C = 0.30(2)$~cm$^3$K/Cr, and $\theta = 4.1(6)$~K\@.  The value of $C$ is smaller than the value $0.375$~cm$^3$K/Cr expected for $S = 1/2$ with a $g$-factor $g = 2$.  However, out of the $7$ Cr ions in each formulae unit of Ca$_{10}$Cr$_7$O$_{28}$, $6$ are in Cr$^{+5}$ valence state with $S = 1/2$ and $1$ Cr ion is in Cr$^{+6}$ valence state and is expected to be non-magnetic with $S = 0$.  Therefore, the value of the Curie constant per magnetic Cr ion will be $C = 7/6 \times$ the value $0.30(2)$~cm$^3$~K/Cr found above.  This gives $C \approx 0.35(2)$ which in turn leads to an effective magnetic moment $\mu_{eff} \approx 1.68(5)~\mu_B$ which is close to the value $1.73~\mu_B$ expected for spin $S = 1/2$ with $g$-factor equal to $2$. 
 
The value of $\theta = 4.1(6)$~K is small and positive indicating weak ferromagnetic exchange interactions.  This is consistent with the huge increase in $\chi$ at low $T$ seen in Fig.~\ref{Fig-chi}~(a).  The magnetic exchange interactions are however, more complex as can be inferred from the deviation below about $T = 20$~K of $1/\chi(T)$ from the Curie-Weiss fit as can be seen in Fig.~\ref{Fig-chi}~(b)~inset.  The $1/\chi(T)$ data deviates upwards of the Curie-Weiss fit which means that the $\chi(T)$ data becomes smaller than expectation from the fit.  This indicates the presence of antiferromagnetic interactions.  Thus the magnetic susceptibility data suggests the presence of both ferromagnetic and antiferromagnetic exchange interactions with the ferromagnetic interactions being the dominant ones leading to a net positive Weiss temperature $\theta \approx 4$~K\@.   

The magnetization $M$ versus magnetic field $H$ data measured at various temperature $T$ are shown in Fig.~\ref{Fig-MH}.  For $T \geq 50$~K, $M(H)$ isotherms are linear in $H$.  For the $M(H)$ data at $T = 25$~K one observes a slight curvature.  However, the $M(H)$ isotherms at $T = 5$ and $1.8$~K show clear curvature with tendency of saturation.  This is consistent with net ferromagnetic interactions which are weak.  The data at $T = 1.8$~K, although still increasing with $H$, are near saturation to values close to $\approx 80$\% of the value expected ($M = 1~\mu_B/$Cr) for $S = 1/2$ moments.  This is again consistent with only 6 out of 7 Chromium ions being magnetic.  The fact that the magnetic moments can be saturated at magnetic fields $\sim 5$~T also suggests that the energy scale of the largest magnetic exchange interactions in Ca$_{10}$Cr$_7$O$_{28}$ is small $\sim 10$~K\@.   

\begin{figure}[t]   
\includegraphics[width= 3 in]{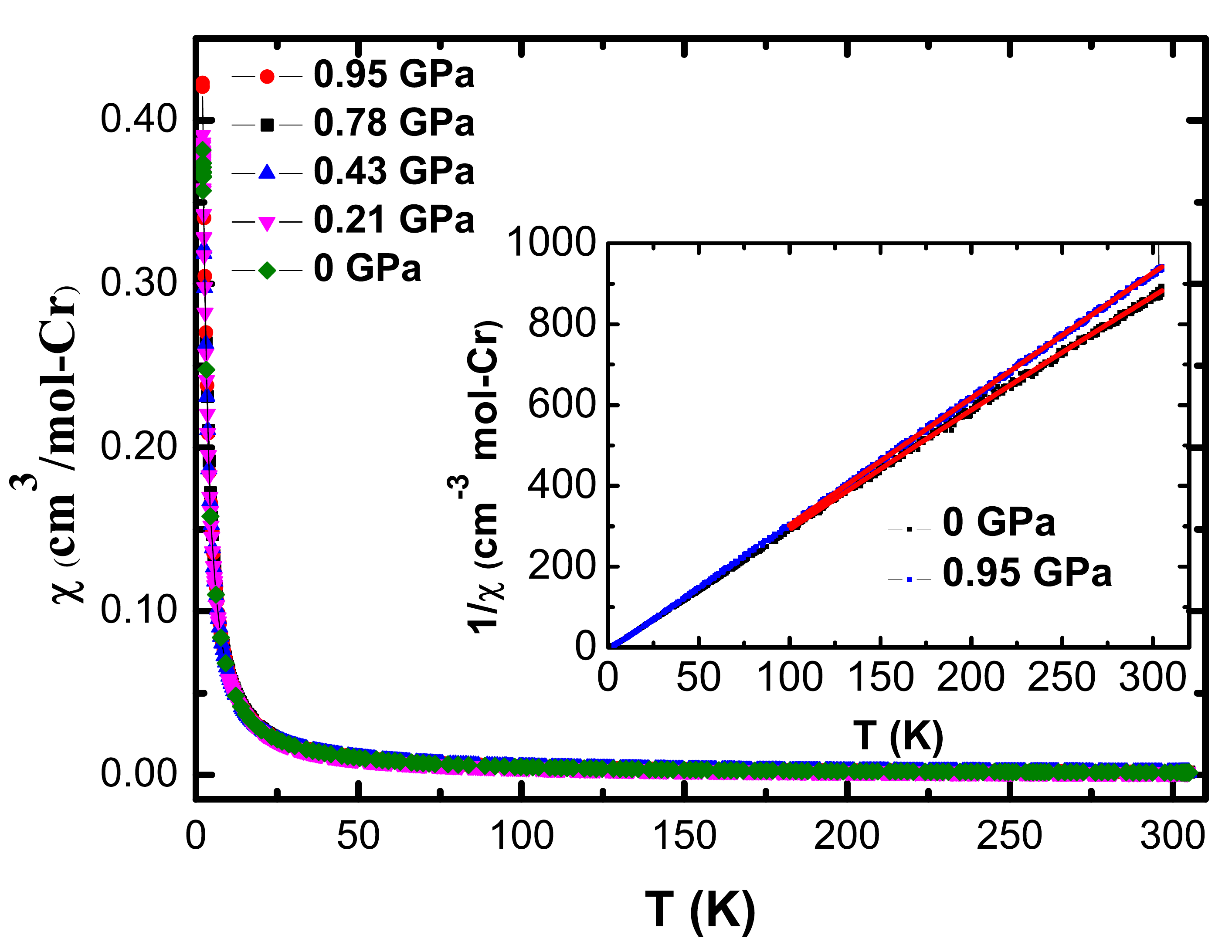}    
\caption{(Color online) Magnetic susceptibility $\chi$ versus $T$ for Ca$_{10}$Cr$_7$O$_{28}$ measured in various externally applied pressures $P$ in a magnetic field $H = 1$~T\@.  The inset shows the $1/\chi(T)$ versus $T$ data for $P = 0$ and $P \approx 1$~GPa.  The solid curve through the data in the inset are fits to a Curie-Weiss expression.  The data is presented per Cr in the formula unit (which is 7).   
\label{Fig-HP-chi}}
\end{figure}

\subsection{High Pressure Magnetic Susceptibility}

If the spin-liquid state in Ca$_{10}$Cr$_7$O$_{28}$ is stabilized by a delicate balance of several magnetic exchanges \cite{Balz2016} then pressurizing the material may disturb this balance and lead to a destruction of the spin-liquid state and may in turn lead to the stabilization of a magnetically ordered state.  Additionally, the Kagome-bilayers, which are magnetically isolated at ambient pressure\cite{Balz2016} may start interacting if brought closer.  With this motivation we have performed high pressure measurements of the magnetic susceptibility $\chi$ versus $T$ at various applied hydrostatic pressures $P$.  Figure~\ref{Fig-HP-chi} shows the high pressure $\chi(T)$ data for Ca$_{10}$Cr$_7$O$_{28}$.  From the main panel we can see that although the magnitude of $\chi$ at the lowest temperature increases slightly, the basic behaviour of $\chi(T)$ doesn't change upto the highest pressures used in our measurements $P \approx 1$~GPa.  For typical transition metal oxides this pressure amounts to a contraction in the unit cell volume of about $1\%$.  This is a large change in the unit-cell size.  The fact that the $\chi(T)$ doesn't show any significant change suggests that the spin-liquid state in Ca$_{10}$Cr$_7$O$_{28}$ is quite robust and does not hinge on some special values of the exchange parameters.  To make a quantitative analysis of the change in $\chi(T)$ we have fit the high temperature data to a Curie-Weiss behaviour.  The $1/\chi(T)$ data for $P = 0$ and $\approx 1$~GPa are shown in the inset in Fig.~\ref{Fig-HP-chi} and the fits to the Curie-Weiss expression are also shown here as solid curves through the data.  In these fits the effective magnetic moment was fixed at its ambient pressure value.  The value of the Weiss temperature changes from $\theta \approx 4$~K at $P = 0$~GPa to $\theta \approx 7$~K at $P = 1$~GPa suggesting an increase in the relative importance of the existing ferromagnetic exchanges.  This is consistent with the increased magnitude of $\chi$ at the lowest temperatures in applied pressures.

\begin{figure}[h]   
\includegraphics[width= 3 in]{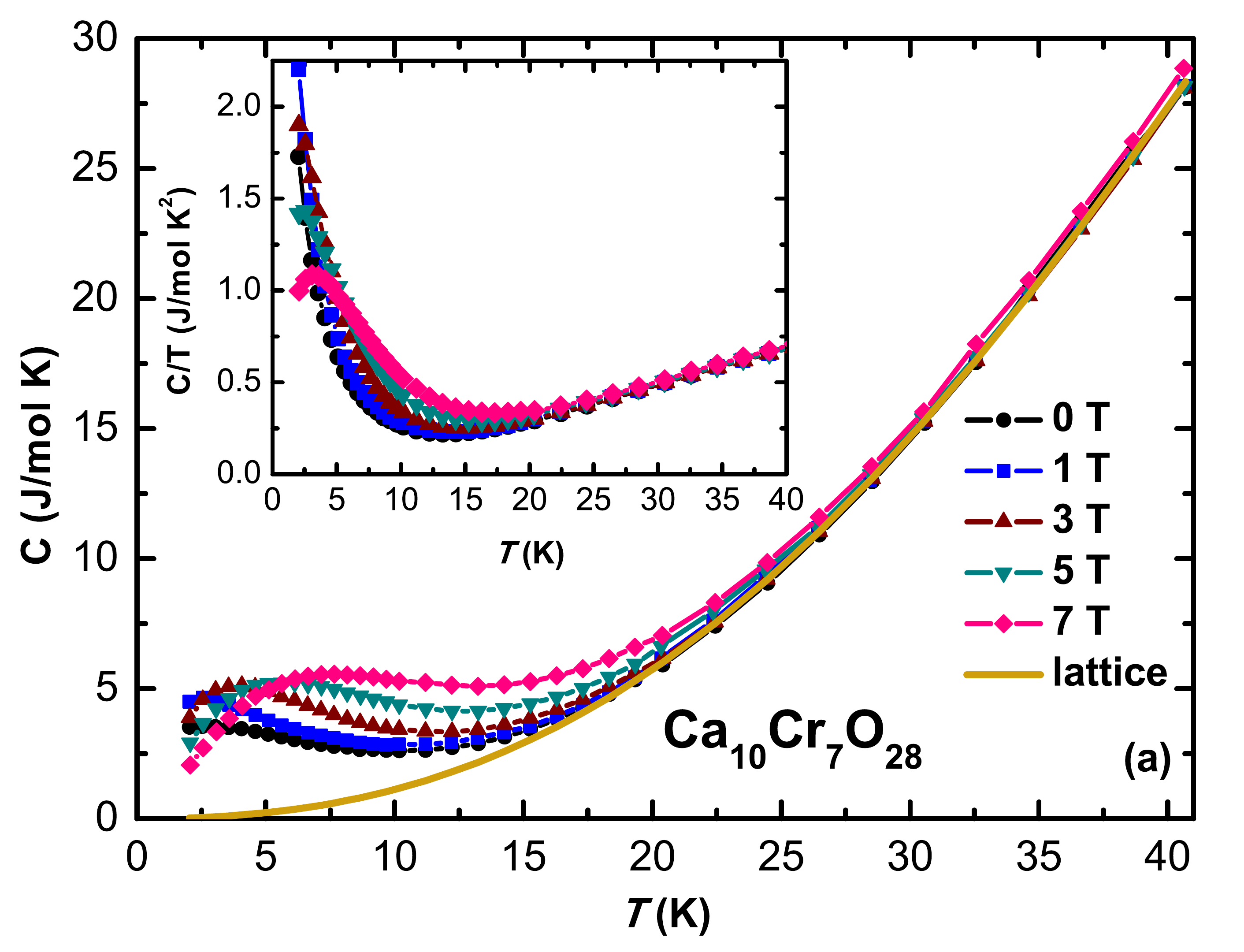}    
\includegraphics[width= 3 in]{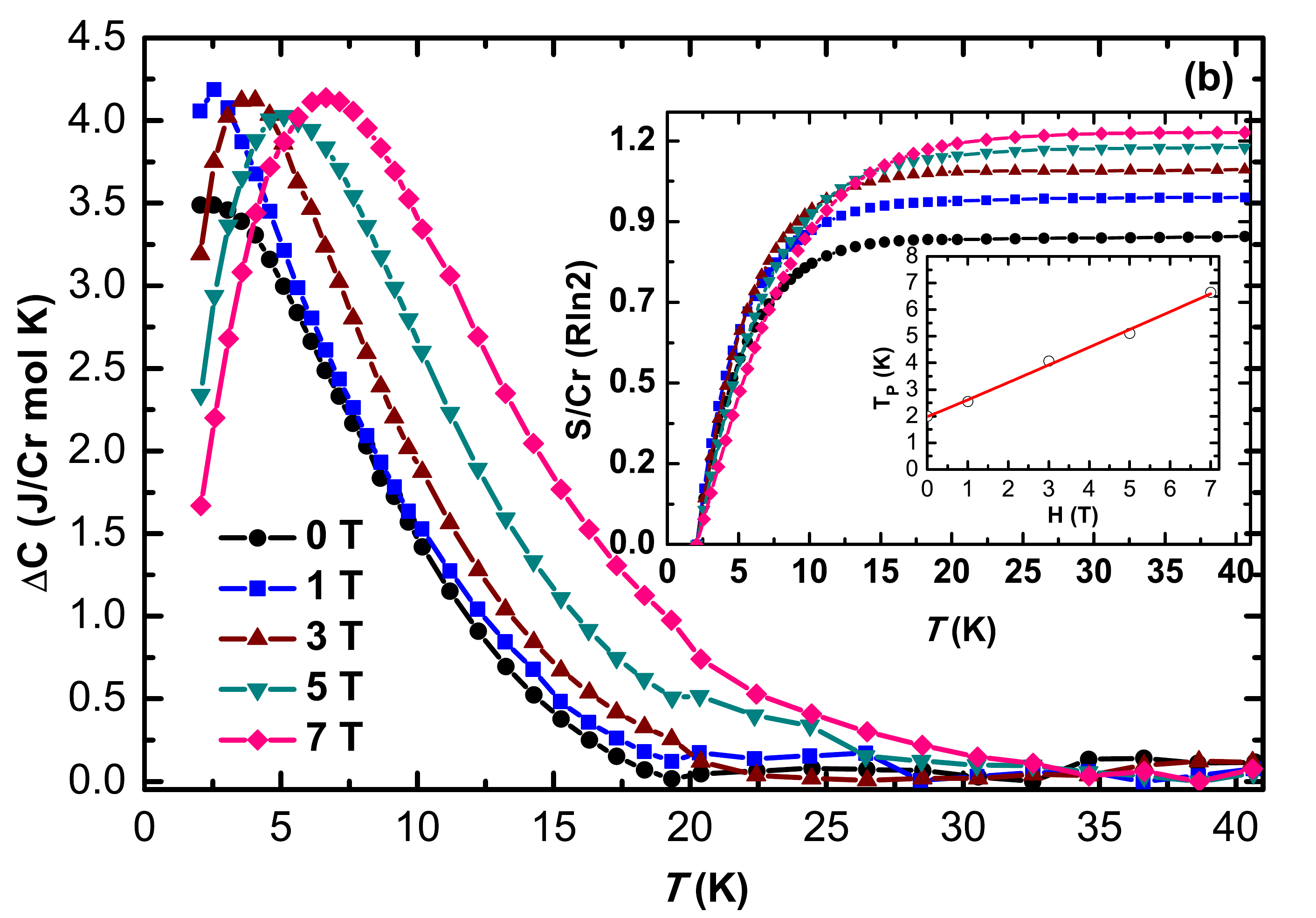}    
\caption{(Color online) (a) Heat capacity $C$ versus temperature $T$ for Ca$_{10}$Cr$_7$O$_{28}$ measured in various magnetic fields $H$.  The inset shows the $C/T$ versus $T$ data.  (b) The difference heat capacity $\Delta C = C -$ lattice versus $T$ for various magnetic fields.  Inset shows the magnetic entropy estimated from the $\Delta C(T)$ data at various $H$.  The inset in the inset shows the peak position $T_P$ versus magnetic field $H$.  The solid curve through the data is a linear fit.   
\label{Fig-Cp}}
\end{figure}
 
\subsection{Heat Capacity}
Heat capacity $C$ versus temperature $T$ data for Ca$_{10}$Cr$_7$O$_{28}$ measured between $T = 1.8$ and $40$~K in various magnetic fields $H$ are shown in the main panel in Fig.~\ref{Fig-Cp}~(a).  The inset shows the same data plotted as $C/T$ versus $T$.  The first thing to note is that the $H = 0$ data below $T = 10$~K shows an upturn and approximately saturates below $T = 3$~K\@.  This might suggest an onset of long-ranged order.  On application of a magnetic field this upturn moves up in temperature and develops into a complete peak at $H = 3$~T\@.  The peak then moves to higher temperatures for larger $H$.  The broad peak is very unlike that expected for a second order phase transition where a $\lambda$-like anomaly is usually seen in the heat capacity.  Thus we believe that this anomaly does not signal a magnetic phase transition.  This is supported by the entropy recovered under the peak as we now describe.  From Fig.~\ref{Fig-Cp}~(a)~inset it is clear that the magnetic fields have a pronounced effect at low temperatures.  However, the data above about $T = 30$~K is independent of $H$ suggesting that it is mostly of non-magnetic origin.  We therefore use the data above $T = 30$~K and extrapolate it to lower temperatures to get an approximate estimation of the lattice heat capacity.    This data is shown as the solid curve in Fig.~\ref{Fig-Cp}~(a).  Using this we can get the magnetic contribution to the heat capacity at various magnetic fields by subtracting the approximate lattice contribution from the $C(T,H)$ data.  This has been done and the resulting difference data $\Delta C(T,H)$ are shown in Fig.~\ref{Fig-Cp}~(b) for all magnetic fields.  The $\Delta C(T,H)$ thus obtained can be used to estimate the magnetic entropy by integrating $\Delta C/T$ versus $T$ data. 
The temperature dependence of the entropy $S(T)$ estimated in this way is shown in Fig.~\ref{Fig-Cp}~(b)~inset for all magnetic fields.  The $S(T)$ data is presented in units of $R$ln2 per magnetic Cr.

We note that $S(T)$ for $H = 0$ reaches about $85\% R$ln2 at $T \approx 30$~K\@.  It is therefore unlikely that a long ranged magnetic order will occur at lower temperatures.  If such a transition does occur it can involve only $15\% R$ln2 entropy which would mean a much reduced moment ordering.  This is consistent with previous $C(T)$ measurements on single crystalline Ca$_{10}$Cr$_7$O$_{28}$ down to $T = 0.3$~K in $H = 0$ that have revealed the absence of long ranged magnetic ordering \cite{Balz2016}. 

In a magnetic field, the anomaly in the magnetic contribution moves to higher temperatures although the magnitude of the peak does not change as can be seen most clearly in Fig.~\ref{Fig-Cp}~(b).  The temperature of the peak $T_P$ as a function of magnetic field $H$ is plotted in the smaller inset in Fig.~\ref{Fig-Cp}~(b).  We find an almost linear dependence of $T_P$ on $H$ as can be seen by the solid curve through the data which is the relation $T_P = 1.986 + 0.63 H$.  The entropy $S(T)$ associated with the peak in the magnetic contribution to $C$, shown in Fig.~\ref{Fig-Cp}~(b)~inset, is also pushed up in temperatures with increasing $H$ and we recover the full $R$ln2 at $H \geq 3$~T\@.  This behaviour of the entropy in a magnetic field is a hallmark of geometrically frustrated magnets.  The frustration suppresses the tendency for long ranged order leading to the accumulation of the entropy of the unordered $S = 1/2$ moments at lower temperatures.  The magnetic field leads to partial alignment of the disordered moments at higher temperatures than at $H = 0$ leading to magnetic entropy being recovered to higher temperatures.   
The observation that fields of $H \leq 7$~T affect the magnetic heat capacity and are able to move the magnetic entropy to higher temperatures again suggests that the magnetic energy scales in Ca$_{10}$Cr$_7$O$_{28}$ are $\sim 10$--$20$~K\@.     

\section{Summary and Discussion:} We have synthesized polycrystalline samples of the recently discovered Kagome bilayer spin liquid material Ca$_{10}$Cr$_7$O$_{28}$ and studied in detail its temperature dependent magnetic susceptibility at ambient and high pressure, isothermal magnetization, and temperature and magnetic field dependent heat capacity measurements.    

The ambient pressure magnetic measurements indicate the presence of both ferromagnetic (FM) and antiferromagnetic (AFM) exchange interactions with the FM interactions dominating.  The net magnetic scale is about $\sim 10$--$15$~K as evidenced by the near saturation of the magnetization at $T = 1.8$~K in a magnetic field of $H = 5$~T\@.  The Curie constant is consistent with one out of the 7 Chromium ions per formula unit being non-magnetic.  The magnetic Cr ions form a bilayer Kagome lattice.  
High pressure magnetic susceptibility measurements up to $P \approx 1$~GPa reveal that the spin-liquid state at ambient pressure is quite robust and may not depend on a delicate balance between any specific values of competing exchange interactions.  Additionally, our results indicate that at high pressure the relative strength of ferromagnetic interactions increases as evidenced by an increase in the value of the Weiss temperature from $\theta = 4$~K at $P = 0$ to $\theta = 7$~K at $P = 1$~GPa.    

The heat capacity in $H = 0$ shows an incomplete anomaly peaked around $\approx 2$~K\@.  The entropy recovered between $T = 1.8$~K and $T = 30$~K in $H = 0$ is close to $85\%$~Rln2 suggesting absence of magnetic ordering for lower temperatures, consistent with a spin-liquid state.    
This broad anomaly peaked in $C$ is consistent with previous zero-field measurements on single crystal Ca$_{10}$Cr$_7$O$_{28}$ and was associated with the onset of coherent quantum fluctuations \cite{Balz2016}.  A similar broad anomaly has been observed for several other QSL candidates.  For example, the organic triangular lattice spin liquid EtMe$_3$Sb[Pd(dmit)$_2$]$_2$,\cite{M-Yamashita2010} shows an anomaly in the heat capacity at $\approx 6$~K while the recently discovered QSL candidate YbMgGaO$_4$ shows a heat capacity anomaly at $\approx 2.4$~K~ \cite{Li2015}.  This anomaly for QSLs is understood to be a crossover from a thermally disordered state to a quantum disordered state.  A low temperature anomaly in the heat capacity which moves up in temperatures on the application of magnetic field is a hallmark of frustrated magnets in general.  For example, in addition to the above materials, the pyrochlore spin-ice material Pr$_2$Zr$_2$O$_7$ also shows an anomaly in the heat capacity at $\approx 2$~K which is attributed to the formation of a collective spin-ice state.  This anomaly has been shown to move approximately linearly to higher temperatures with magnetic fields \cite{Petit2016}. 

The heat capacity anomaly for Ca$_{10}$Cr$_7$O$_{28}$ occurs at $2.4$~K at $H = 0$ and moves to higher temperatures in a magnetic field approximately linearly.  Since the fields of our measurements are much smaller than the saturation fields of $12$--$13$~T \cite{Balz2016, Balz2016a}, the linear dependence of the peak temperature with $H$ is intriguing.  Since the ferromagnetic exchange is dominant in Ca$_{10}$Cr$_7$O$_{28}$ it is possible that in-plane short-ranged order develops and is strengthened in a field.  However, future measurements would be needed to understand this observation.

\noindent
\emph{Acknowledgments.--} We thank the X-ray facility at IISER Mohali.  YS acknowledges DST, India for support through Ramanujan Grant \#SR/S2/RJN-76/2010 and through DST grant \#SB/S2/CMP-001/2013.

\end{document}